\documentclass[traditabstract]{aa}
\usepackage{txfonts}
\usepackage{natbib}
\usepackage{graphicx}
\usepackage{subfigure}
\usepackage{times}
\bibpunct{(}{)}{;}{a}{}{,}
\newcommand{\but}{BU~Tau }
\newcommand{\bue}{BU~Tau}

\newcommand{\ha}{H$\alpha$ }

\newcommand{\ond}{Ond\v{r}ejov }
\newcommand{\dao}{DAO }
\newcommand{\roz}{Rozhen }

\newcommand{\elo}{OHP }
\newcommand{\abs}{absorption }
\newcommand{\emi}{emission }

\newcommand{\Emi}{Emission }

\newcommand{\p}{$\pm$}

\newcommand{\kms}{km~s$^{-1}$ }
\newcommand{\ks}{km~s$^{-1}$}

\newcommand{\ms}{M$_{\odot}$}
\newcommand{\rs}{R$_{\odot}$}

\newcommand{\ANG}{\accent'27A}
\newcommand{\Am}{\ANG~mm$^{-1}$ }

\newcommand{\hae}{H$\alpha$}

\newcommand{\he}{\ion{He}{i}~6678~\ANG\ }

\begin{document}

   \title{Properties and nature of Be stars
\thanks{Based on new spectral and photometric observations from
 the following observatories: Dominion Astrophysical Observatory, Herzberg Institute of Astrophysics, National Research Council of Canada,
 Haute Provence, IGeoE-Lisbon, Astronomical Institute AS CR Ond\v{r}ejov, and Rozhen.}
\subtitle{27. Orbital and recent long-term variations of the Pleiades Be star\\
                              Pleione = BU~Tauri}}

\author{J.~Nemravov\'a\inst{1}\and P.~Harmanec\inst{1}\and
  J.~Kub\'at\inst{2}\and P.~Koubsk\'y\inst{2}\and L.~Iliev\inst{3}\and
   S.~Yang\inst{4}\and J.~Ribeiro\inst{5}\and M.~\v{S}lechta\inst{2}\and
   L.~Kotkov\'a\inst{2}\and M.~Wolf\inst{1}\and P.~\v{S}koda\inst{2}}

   \institute{
    Astronomical Institute of the Charles University,
    Faculty of Mathematics and Physics,\\
    V Hole\v sovi\v ck\'ach 2, CZ-180 00 Praha 8, Czech Republic
\and
    Astronomical Institute of the Academy of Sciences,
    CZ-251~65~Ond\v{r}ejov, Czech Republic
\and
   Institute of Astronomy, Bulgarian Academy of Sciences, BG-1784,
   72 Tsarigradsko Chaussee Blvd., Sofia, Bulgaria
\and
   Physics \& Astronomy Department, University of Victoria,
   PO Box 3055 STN CSC, Victoria, BC, V8W 3P6, Canada
\and
   Observat\'orio do Instituto Geogr\'afico do Ex\'ercito,
   R.~Venezuela 29, 3 Esq. 1500-618, Lisboa, Portugal
  }

   \titlerunning{Duplicity of \bue}

  \offprints{J.~Nemravov\'a,\\ \email: janicka.ari@seznam.cz}

   \date{Release \today}

   \date{Release \today}

\abstract{Radial-velocity variations of the \ha emission measured
on the steep wings of the \ha line, prewhitened for the long-time
changes, vary periodically with a period of 218\fd025\p0\fd022, confirming
the suspected binary nature of the bright Be star \bue,
a~member of the Pleiades cluster. The orbit seems to have
a high eccentricity over 0.7, but we also briefly discuss the possibility
that the true orbit is circular and that the eccentricity is spurious owing
to the phase-dependent effects of the circumstellar matter. The projected
angular separation of the spectroscopic orbit is large enough to allow
the detection of the binary with large optical interferometers, provided
the magnitude difference primary $-$ secondary is not too large.
Since our data cover the onset of a new shell phase up to development
of a metallic shell spectrum, we also briefly discuss the recent long-term
changes. We confirm the formation of a new envelope, coexisting with
the previous one, at the onset of the new shell phase. We find that
the full width at half maximum of the \ha profile has been decreasing
with time for both envelopes. In this connection, we briefly discuss
Hirata's hypothesis of precessing gaseous disk
and possible alternative scenarios of the observed long-term changes.
\keywords{stars: early-type -- stars: binaries -- stars: Be --
          stars: individual: \bue}
}

\maketitle

\section{Introduction\label{1}}
Pleione (\bue, 28~Tau, HD~23862) is a well-known Be star and a member of
the Pleiades cluster. It underwent several phase transitions between
B, Be, and Be shell phases, accompanied by pronounced light variations;
see, e.g. \citet{gul77}, \citet{sharov76}, \citet{ili88}, \citet{sharov92},
\citet{hirata76}, \citet{hirata77}, \citet{hirata95}, \citet{doazan88},
\citet{ili2007}, and \citet{tana2007}.

There is a rather complicated history of attempts to study
the radial-velocity (RV hereafter) variations of this star. \citet{struve43}
measured RVs on the photographic spectra taken in the years 1938-1943  and
tentatively concluded that the RV of \but varies with a possible period
of 142~days or -- less likely -- 106 days. \citet{mer52} studied RVs
from 1941 to 1951 and found clear long-term variations with some
overlapping changes on a shorter time scale.  \cite{gul77} analyzed
a large collection of digitized photographic spectra from 1938-1954
and from 1969-1975 and concluded that there are no significant RV
changes. \citet{bal88} carried out an analysis of a homogeneous series
of Haute Provence high-dispersion photographic spectra from 1978-1987
and once more concluded that the shell RVs vary with periods of 136.0
and 106.7 days. \citet{kat96a,kat96b} analyzed shell RVs from the two
consecutive shell phases separated some 34 years, using published as well
as new RVs and concluded that \but is a spectroscopic binary with
an orbital period of 218\fd0, semi-amplitude of 5.9~\ks, and a large
orbital eccentricity of 0.60. However, \citet{rivi2006} -- analyzing
a series of electronic spectra -- were unable to confirm the 218-d
period and concluded that \but is not a spectroscopic binary.
\citet{hirata2007} analyzed a long series of polarimetric observations
and presented a model of a slowly precessing disk to explain the long-term
B -- Be -- Be shell phase transition. He argued that the disk
precession is caused by the attractive force of the secondary in the 218-d
binary.
\citet{bebin82} compiled the majority of at that time available RVs
of \but and averaged them over about 100 days. This resulted in
a smooth RV curve with a period of about 13000 days (35.6 years),
in phase with the recorded shell episodes. \citet{bebin82} speculated
that \but could be a long-periodic binary with shell phases occurring
always at the same orbital phases. A more distant companion with
an angular distance of 0\farcs22 was indeed discovered from speckle
interferometry by \citet{mcal89}. \citet{gies90}
studied a sequence of low-dispersion \ha spectra of \but taken with
a sampling rate of 7~ms during a lunar occultation on 1987~March~6.
They detected an asymmetry of the envelope in agreement with the observed
long-term $V/R$ changes. They speculated that the speckle-interferometric
component could have an eccentric orbit and that the recurrent shell phases
could be caused by its periastron passages. \citet{luth94}
compiled RVs from the years 1938-1990 and confirmed a period of
12450-12860~days. They advocated an eccentric orbit and mass transfer
resulting in a release of a new shell during periastron passages, but
the gaps in their RV curve do not allow one to conclude that the orbit has
a high eccentricity. Finally, using the technique of adaptive optics
photometry and astrometry, \citet{rob2007} report discovery of a new
companion to \but at a separation of 4\farcs66 with a spectral type M5.
They also confirm a companion at 0\farcs24 and discuss other suggested
companions.

\begin{figure}
\centering
\resizebox{\hsize}{!}{\includegraphics{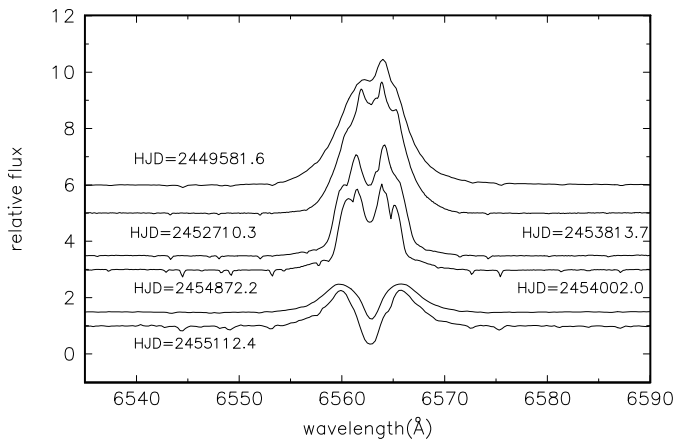}}
\caption{Comparison of \ha profiles from different stages of the long-term
changes.}\label{profile}
\end{figure}

\begin{figure}
\centering
\includegraphics[width=\hsize]{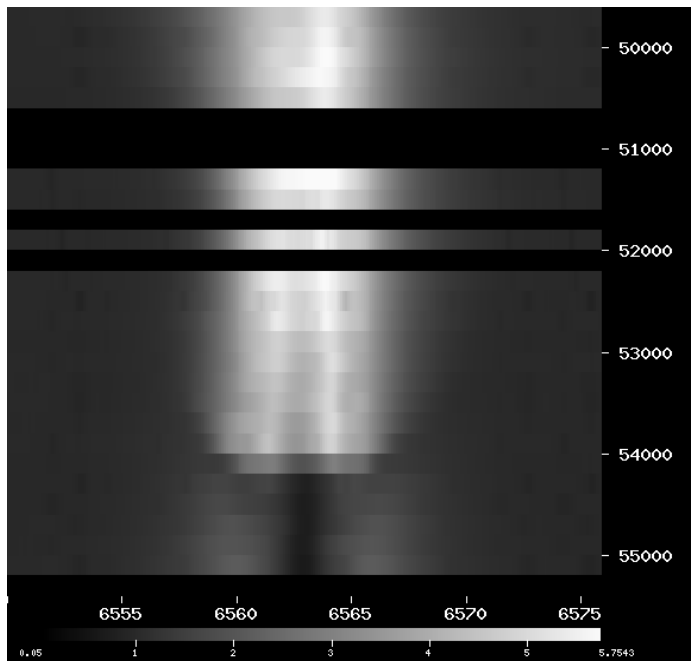}
\caption{A complete series of our \ha profiles in a gray representation
(only a few saturated or underexposed spectra were omitted). Abscissa shows
the wavelength scale in \ANG, while the time on ordinate is shown
in JD-2400000. Each horizontal strip represents an average of spectra
secured within 200 days, and dark horizontal belts correspond to time
intervals from which no spectra are available. At the bottom, there is a scale
showing the correspondence between the flux level in the units of continuum
and the gray scale.}
\label{h3time}
\end{figure}

\begin{table}[h]
\begin{flushleft}
\caption[]{Journal of new spectroscopic observations for \bue.}\label{jou}
\begin{tabular}{rcrccc}
\hline\noalign{\smallskip}
Station&Time interval  &No. of. &Wavelength \\
Source &(HJD$-$2400000)&obs.    &region (\ANG)\\
\noalign{\smallskip}\hline
\hline\noalign{\smallskip}
 \ond & 49581 - 54872      & 101    & 6200 - 6800\\
 \dao & 49786 - 54912      &  26    & 6150 - 6700\\
 \elo & 51569 - 52664      &  21    & 6200 - 6700\\
 \roz & 52710 - 54108      &  23    & 6520 - 6610\\
 Lisboa & 54874 - 54881      &   4    & 6520 - 6600\\
\noalign{\smallskip}\hline
\end{tabular}
\end{flushleft}
\end{table}

  We succeeded in collecting a rich series of electronic spectra
at several observatories, covering many cycles of the suspected 218-d
period. The main goal of this study is, therefore, to resolve the issue
of whether \but is a spectroscopic binary. \citet{kat96a,kat96b}
based their orbit on the RV measurements
of shell lines that may be affected by possible asymmetries in
the circumstellar matter. Moreover, their RV curve has a rather small amplitude
and is based on a collection of heterogeneous data. It naturally shows
a rather large scatter around the mean curve.
The spectra at our disposal all cover the red spectral
region near \hae. They were taken over the time interval when
the star had fairly strong \ha emission. Therefore, our study is based
on the RV measurements of the steep wings of the emission, which
is a procedure that turned out to be successful for detecting the
duplicity of several other Be stars
\citep{bozic95,zarf19,zarf20,miro2001,miro2002}.

\begin{figure}
\centering
\resizebox{\hsize}{!}{\includegraphics{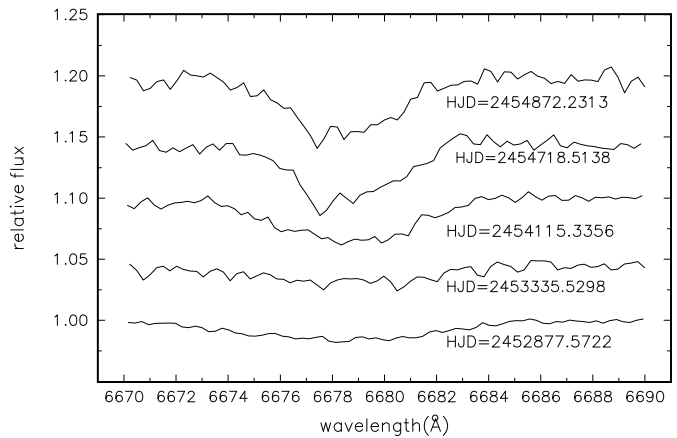}}
\caption{Selected \he line profiles, ordered in time,
with corresponding HJDs.}\label{heexam}
\end{figure}

 Since very pronounced long-term spectral variations occurred over
the time interval covered by our spectra, we also briefly describe
these changes and discuss them, especially in relation to the model
put forward by \citet{hirata2007}.

\begin{figure}
\centering
\resizebox{\hsize}{!}{\includegraphics{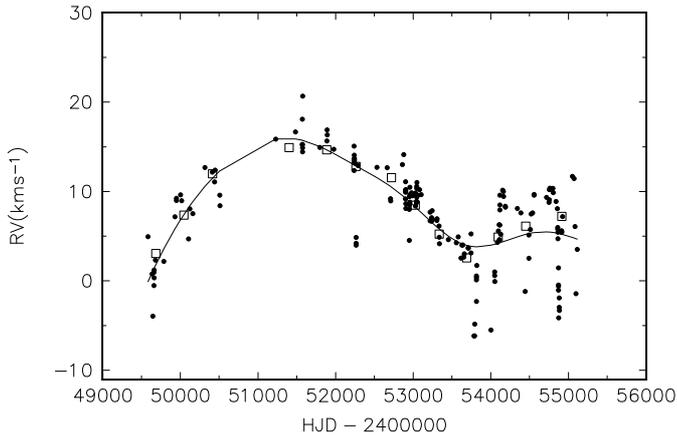}}
\caption{Measured RVs of the \ha emission wings plotted vs. time.
Prewhitening for the long-term changes, carried out with the help
of the program HEC13, is shown by a line. Empty squares show the alternate
way to remove long-term RV changes via individual $\gamma$
velocities for subsets spanning no more than a year.
See the text for details.}
\label{etime}
\end{figure}

\section{Spectroscopic observations and their reductions}
The red spectra at our disposal were obtained at five observatories
and their overview is in Table~\ref{jou}.
Details about the instruments and data reduction can be found in
Appendix~\ref{spered} where also Table~\ref{hjdrv} with our RV measurements
of the steep wings of the \ha emission and of the \ha absorption core
is provided. The latter was measured for comparison with the RVs collected
and analyzed by \citet{kat96b}, but only for those spectra where the
absorption was clearly visible.

\begin{figure}
\centering
\resizebox{\hsize}{!}{\includegraphics{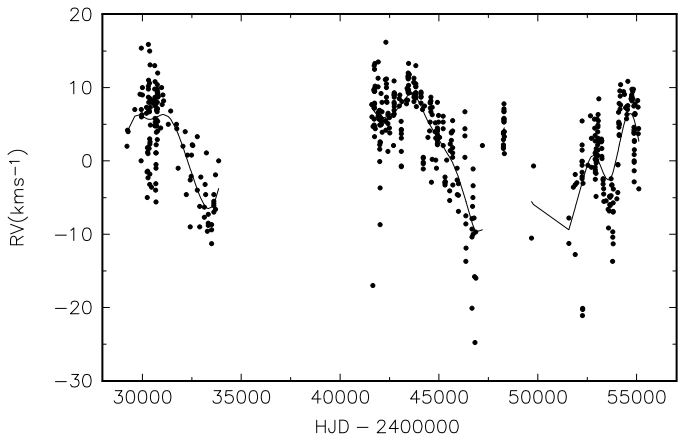}}
\caption{Measured \ha absorption core plotted vs. time. We also
included the RV measurements of shell lines
by \citet{kat96b} and \citet{rivi2006}
to this plot. Prewhitening for the long-term changes, carried out with
the help of the program HEC13, is shown by a line. See the text for details.}
\label{atime}
\end{figure}

\begin{figure}
\centering
\resizebox{\hsize}{!}{\includegraphics{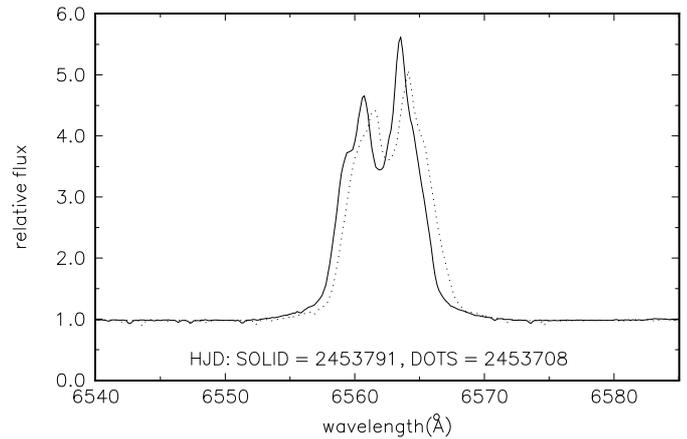}}
\resizebox{\hsize}{!}{\includegraphics{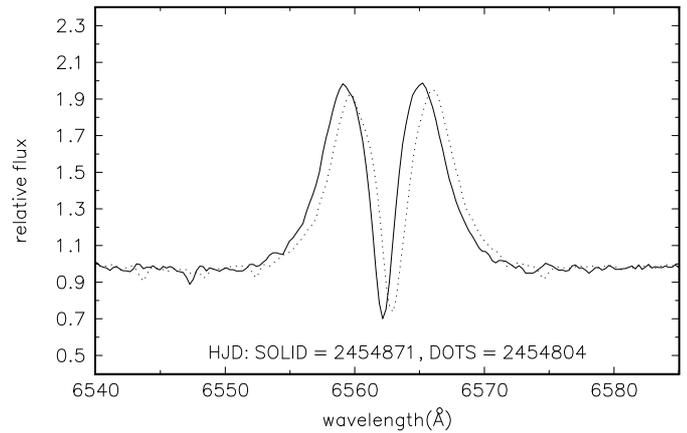}}
\caption{A comparison of two pairs of the \ha line profiles from the
locally recorded velocity extrema (HJDs of the profiles are indicated).}
\label{shift}
\end{figure}

\begin{figure}
\centering
\resizebox{\hsize}{!}{\includegraphics{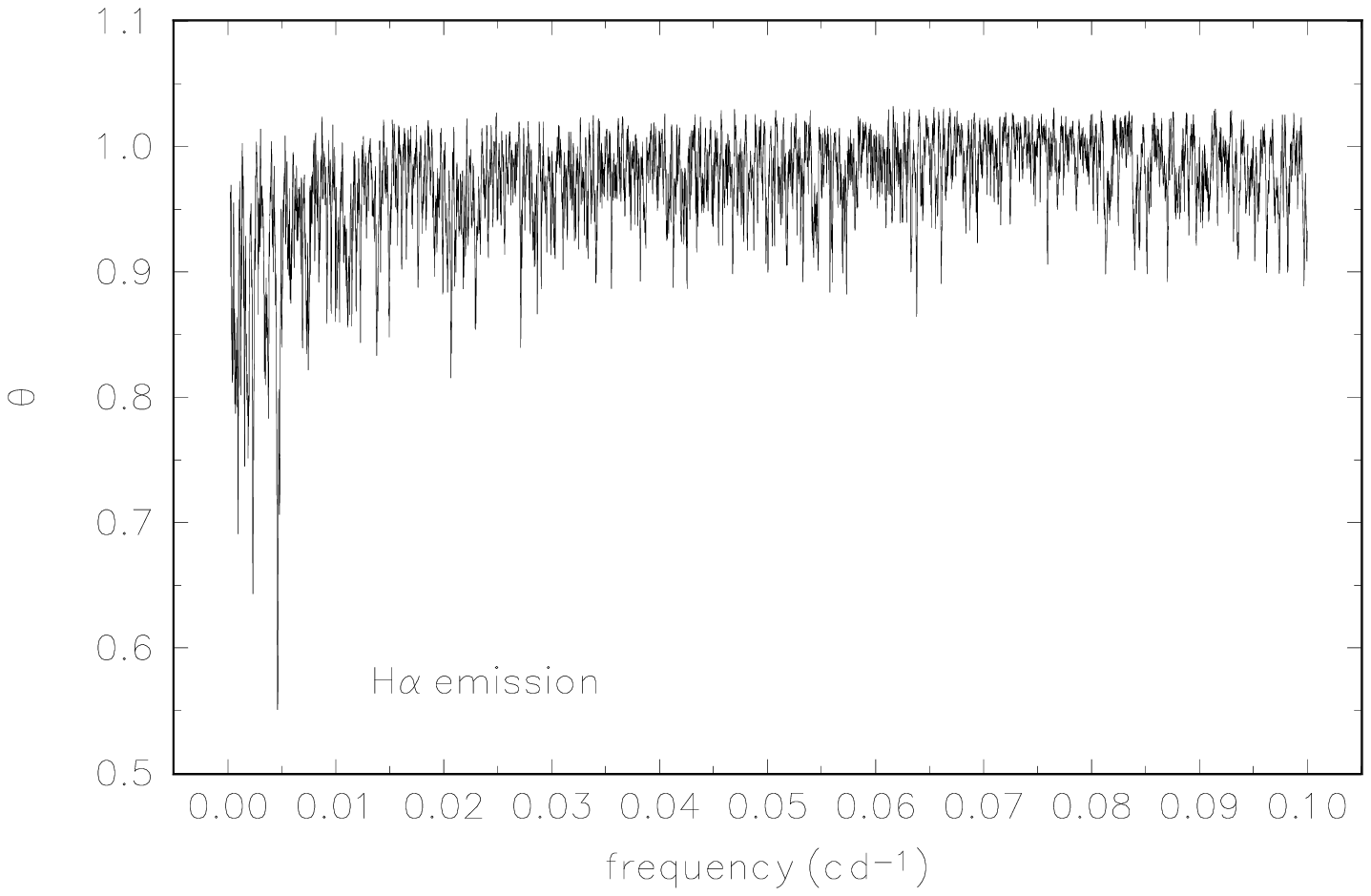}}
\resizebox{\hsize}{!}{\includegraphics{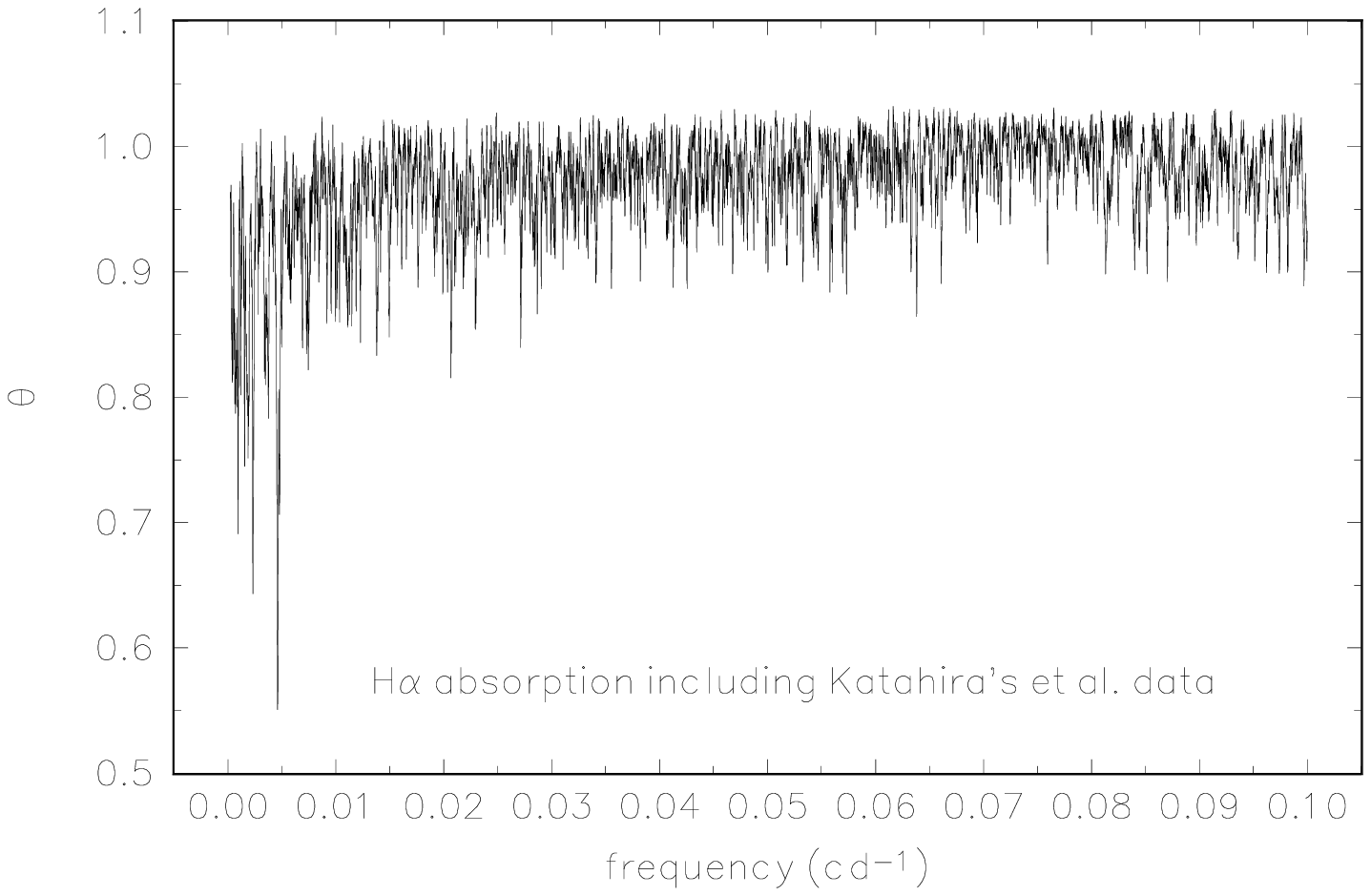}}
\caption{\citet{stel78} PDM $\theta$ statistics for all emission-wing
RVs (top) and shell absorption-core RVs including \citet{kat96b}. The dominant
frequency of 0.004587~c\,d$^{-1}$ corresponds to the 218-d period.}
\label{theta}
\end{figure}

Over the interval of the more than 5000 days covered by our observations,
the strength of the \ha emission gradually declined and the shape
of the \ha profile underwent notable changes. Typical examples
for several distinct stages are shown in Fig.~\ref{profile}, and the whole
development of a new shell and metallic-shell phase is shown as a
gray-scale representation of all usable \ha profiles in Fig.~\ref{h3time}.

\begin{figure}
\centering
\rotatebox{-180}{\resizebox{\hsize}{!}{\includegraphics{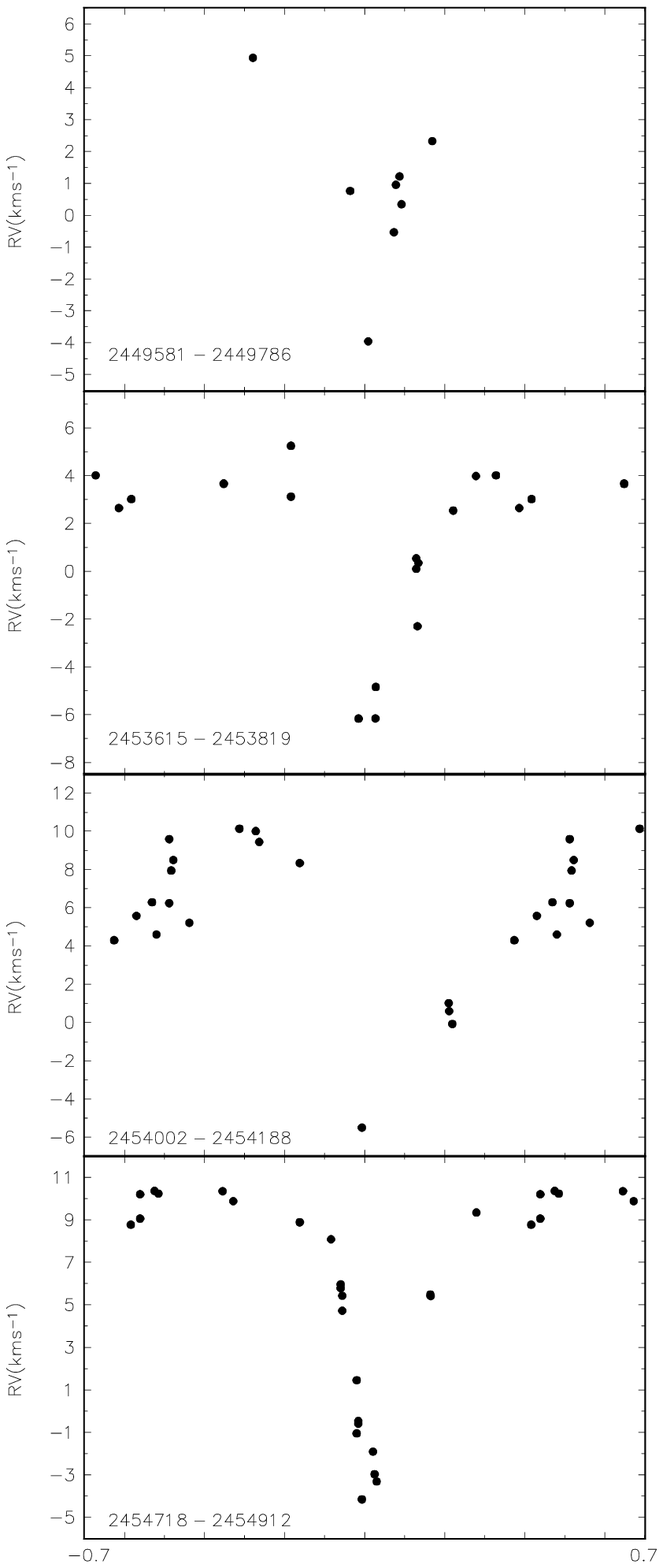}}}
\caption{Orbital RV curves of the \ha emission shown for subsets
of data spanning less than a year. For all plots, period 218\fd053
was used, with phase zero at HJD~2452041.11, which corresponds to
the RV minimum (see Table~\ref{elem13}).}\label{rwesubsets}
\end{figure}

The fading of the \ha emission was accompanied by a light decrease
in the $J$, $H$, $K$, and $L$ IR photometric bands that started
around JD~2451500 \citep{tara2008}. This clearly corresponds to
the gradual development of the hydrogen shell spectrum according to our
spectra -- see also \citet{tana2007}.
 \Emi has been slowly fading from JD~2453000 until now, when its peak intensity
represents only about 30\% of the intensity seen in our earliest
spectra. During the transition from a single-peaked to double-peaked emission,
there is some time interval when the \ha profile has
a characteristic wine-bottle shape.
The occasional presence of additional absorption components has been already noted by
\citet{ili2007} or \citet{tana2007} and is typical of all recorded
shell phases of \bue. Besides the occasional presence of one or more additional
absorptions, extended red emission wings are seen on some \ha profiles.
This makes the emission wings asymmetric and hard to measure for RV.
We also note that all double-peaked profiles recorded prior to
about JD~24540000 always have a red peak stronger than the violet one.
Figure~\ref{h3time} shows that the {\sl width} of the \ha emission has remained
more or less constant over the whole time interval covered by our observations.
The same figure also shows that the metallic shell phase appeared rather abruptly.

Figure~\ref{heexam} shows the gradual development of the \he line profile.
It illustrates well how shallow the line is at the beginning of a new
shell phase. A very interesting finding is that, even for the B8 star,
a presumably photospheric \ion{He}{I} line can develop a shell component.
The profile clearly gets stronger and narrower as the hydrogen shell line
gets deeper. The additional absorption at the blue wing of the line
seen on more recent spectra is the \ion{Fe}{II}~6677.305~\ANG\ shell line.

\section{Radial-velocity changes}\label{rvvar}
Figures~\ref{etime} and \ref{atime} are the time plots of the measured RVs
vs. time for the \ha emission wings and the absorption core. In the later,
we also included all shell RVs used and published by
\citet{kat96b} and \citet{rivi2006}.
One can see systematic RV changes on at least two distinct time scales:
a smooth change on a longer time scale and overlapping more rapid changes,
especially the occasional steep decreases in RV.

\begin{figure}
\centering
\rotatebox{-180}{\includegraphics[width=\hsize]{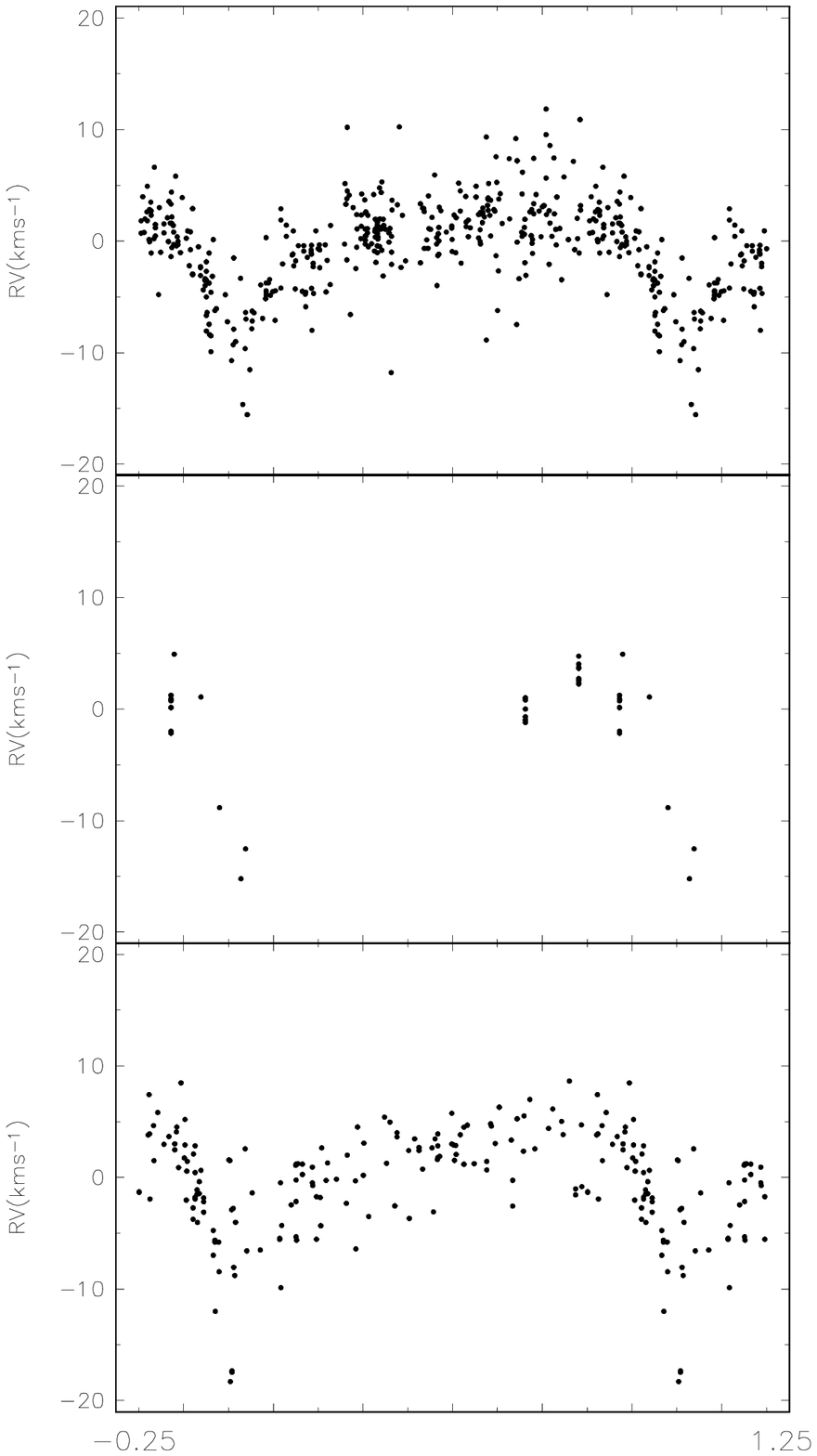}}
\caption{{\sl Top:} The phase plots of all available
Balmer absorption RVs, prewhitened for the long-term RV variations
with HEC13 (as shown in Fig.~\ref{atime}).
Elements from solution~1 of Table~\ref{elem13} were used, with phase
zero at minimum RV. For clarity, we show three different data subsets
separately: {\sl Top panel:} Photographic RVs from \citet{kat96b};
{\sl Central panel:} RVs from electronic Heros spectra published
by \citet{rivi2006}; {\sl Bottom panel:} RVs from electronic spectra
used in this paper.}\label{abs218}
\end{figure}

Considering the uncertainties in accurate RV measurements
combined with the fact that the full amplitude of the changes
is low, it was deemed useful
to convince readers that the RV changes are not only a result of
changing asymmetry of the profiles, but they also represent a real shift of
the whole line. To this end, we compare in Fig.~\ref{shift}
two pairs of the \ha line profiles obtained near the local RV extrema.
The upper pair comes from the beginning of a new shell phase and the bottom
one from a more recent time when a weaker emission and deeper shell cores
are present in the profiles (note a large difference in the
flux scale of the two plots).
The RV shift of the whole emission and absorption core is seen beyond any doubt.
We, therefore, conclude that our RV measurements reflect
{\sl real RV variations} of \bue.

In accordance with \citet{kat96b}, we find that the evolution of the emission
episode is accompanied by long-term RV changes that need to be
removed prior to a search for possible periodic RV changes. To also make
this step as objective as possible, we used two different procedures.

One is that we smoothed the long-term changes using the program HEC13,
written by PH and based on a smoothing technique developed
by \citet{vondrak69,vondrak77}.\footnote{The program HEC13
with brief instructions how to use it is available to interested users at
{\sl http://astro.troja.mff.cuni.cz/ftp/hec/HEC13}\,.}
For both emission and absorption RVs, optimal smoothings were obtained
for the smoothing parameter $\varepsilon=10^{-16}$ fitted through 200-d
normals. (Inspecting the time plots of RVs, we identified $\sim 200$
days as a time scale on which more rapid changes were observed, and this
was the reason for the choice of 200-d normals. We have verified,
however, that the result of smoothing is not sensitive to the
particular choice of the averaging interval for the smoothing within
reasonable limits.)
The RV residuals from the smoothing were subjected to a period search
based on the \citet{stel78} PDM technique over a period range
from 5000 down to 0.05~d. The dominant frequency found in both searches
was 0.004587~c\,d$^{-1}$ and its integer submultiples. The one-day aliases
were largely supressed thanks to having data from observatories, that have
a large difference in their local time,
producing much shallower minima in the $\theta$
statistics ($\sim 0.75-0.82$) and scattered phase diagrams.
To make the diagrams readable, we show the corresponding
$\theta$ statistics in Fig.~\ref{theta} for the emission (top) and
absorption (bottom) RVs only for a limited frequency interval down to
0.1~c\,d$^{-1}$. The result seems to confirm the 218-d periodicity
discovered by \citet{kat96b}.

As another demonstration that the 218-d period is real,
we show phase plots in Fig.~\ref{rwesubsets} for
the original RVs (without prewhitening for the long-term
changes) for several subsets of data covering time intervals no longer than
one year. Clearly similar RV curves, with sharp minima, rather flat
maxima, and a mutual phase coherence, are seen in all cases. The first
subset is based solely on the RVs from the Ond\v{r}ejov spectra secured
with Reticon detector, which were already investigated by \citet{rivi2006}.

\section{\but as a spectroscopic binary}\label{elements}

\begin{figure}
\rotatebox{-180}{\resizebox{\hsize}{!}{\includegraphics{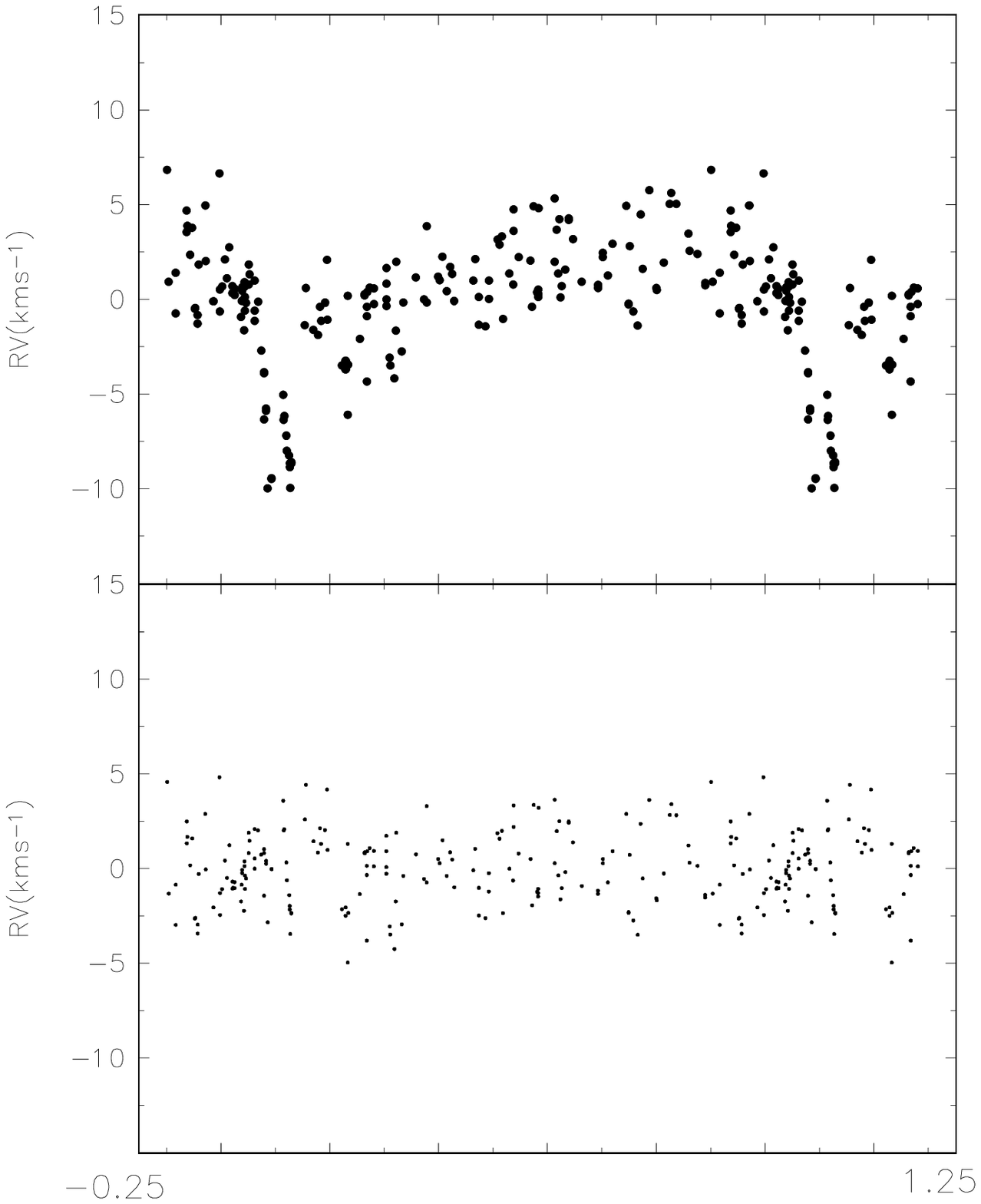}}}
\rotatebox{-180}{\resizebox{\hsize}{!}{\includegraphics{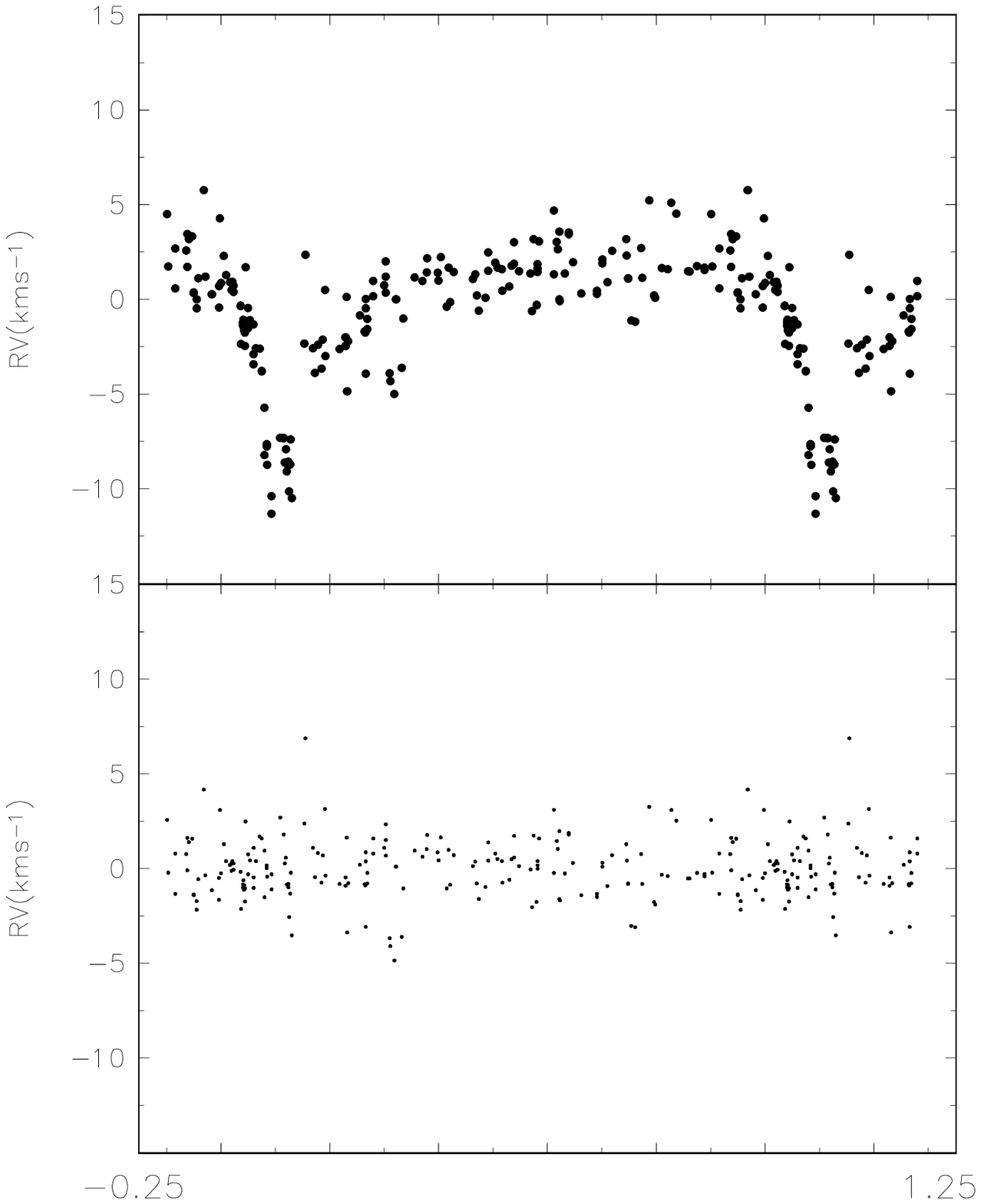}}}
\caption{The orbital RV curves of \but based on the \ha emission RVs
plotted for the solutions~2 and 3 of Table~\ref{elem13}.
Phase zero corresponds to the respective epoch of minimum RV and the O-C
deviations from the solutions are shown by small circles in separate
panels. {\sl Top two panels:} RVs prewhitened via HEC13 (solution~2);
{\sl Two bottom panels:} original RVs minus locally
derived systemic $\gamma$ RVs (solution~3).
See the text for details.}\label{em218}
\end{figure}

Our findings, and especially the fact that the \ha Balmer
emission line moves in RV as a whole in spite of very large secular
changes of its strength, indicate that \but is indeed a single-line
spectroscopic binary that moves in a highly eccentric orbit.
We therefore used the program SPEL (written by the late Dr.~Ji\v{r}\'\i\ Horn
and never published) to derive the orbital elements.
For comparison with \citet{kat96b}, we first derived orbital elements
for Balmer {\sl absorption} RVs, using the data from their study,
RVs published by \citet{rivi2006}, and our own \ha absorption RVs, prewhitened
with HEC13 as shown in Fig.~\ref{atime}. The resulting orbital
elements are given as solution~1 in Table~\ref{elem13} and the corresponding
phase plots are shown in Fig.~\ref{abs218}. For more clarity, we plot there
the photographic RVs, Heros RVs from \citet{rivi2006}, and our \ha absorption
RVs in three separate panels. Although \citet{rivi2006} write
that the suspected binary nature of \but could not be confirmed on the basis
of their data, their RVs also nicely follow the 218-d period. This constitutes
yet another support for the reality of this period. Our solution~1 agrees well
with the result of \citet{kat96b}.

Next, we analyzed the emission RVs that we consider as most realistically
describing the true orbital motion. To see how sensitive
the result is to the manner of prewhitening the data we derived the elements
not only for the RVs prewhitened with the help of HEC13 (see above) but
also from the original data. To this end, we divided the data into
subsets spanning no more than one year and allowed SPEL to derive separate
$\gamma$ velocities for individual data subsets.
The results are summarized in Table~\ref{elem13}, and
the corresponding RV curves compared in Fig.~\ref{em218}.

\begin{table}[h]
\begin{flushleft}
\caption{Several sets of orbital elements:
Solution 1... Katahira, Rivinius and this paper, prewhitened with HEC13;
Solution~2... New emission-line RVs prewhitened with HEC13;
Solution~3... New emission-line RVs with allowance for locally derived
$\gamma$~velocities.}\label{elem13}
\begin{tabular}{llllccc}
\hline\noalign{\smallskip}
 Solution:             & 1                & 2             & 3         \\
Orbital                & Old \& new       & \ha emis.     & \ha emis. \\
element                & \ha abs.         & wings         & wings      \\
\noalign{\smallskip}\hline
\hline\noalign{\smallskip}
$P$ (d)                & 218.023\p0.023   &218.099\p0.050 & 218.053\p0.053\\
$T_{\rm periastr.}$ (d)& 40040.4\p1.6     &52039.34\p0.69 & 52039.73\p0.73\\
$T_{\rm super.c.}$ (d) & 40032.3          &52035.50       & 52034.79      \\
$T_{\rm min.\,RV}$ (d) & 40044.5          &52040.64       & 52041.11      \\
$e$                    & 0.596\p0.035     &0.774\p0.028   & 0.745\p0.026  \\
$\omega (^\circ)$      & 147.7\p4.5       &154.2\p4.0     & 157.3\p3.5    \\
$K_1$ (\ks)            & 5.41\p0.35       &6.30\p0.63     & 6.39\p0.46    \\
$\gamma$ (\ks)         & -0.15\p0.1       &0.35\p0.15     & --            \\
rms (\ks)              &  3.21            &1.93           & 1.58         \\
\hline\noalign{\smallskip}
\end{tabular}
\end{flushleft}
\end{table}

The inspection of Fig.~\ref{em218} shows that even the \ha emission-wing RVs
are indeed indicative of an orbit with high eccentricity but that there is
also an alternative possibility that the observed deep RV minimum could
be a consequence of some unspecified effect of circumstellar matter,
reminiscent of ``an inverse rotational or Rossiter effect". In this case,
the true orbit could essentially be circular. To this end, we derived
yet another, a circular-orbit solution for the \ha emission RVs prewhitened
for long-term changes via HEC13, omitting all RVs from the phase interval
around phase zero with the most negative RVs. This resulted in
the following elements: $P=218\fd34\pm0.62$,
$T_{\rm super.c.}={\rm HJD}~2452009.9\pm4.8$, $K_1=1.72\pm0.21$~\ks.

Using the eliptical-orbit elements for the \ha emission RVs
from Table~\ref{elem13}, we estimated the basic properties of the binary
from the mass function $f(m)=0.00165$~\ms\ for several plausible orbital
inclinations, assuming a normal mass of the primary corresponding to
its spectral type after \citet{mr88} to be $M_1=2.9$~\ms.

\begin{table}[h]
\begin{flushleft}
\caption{Basic physical properties of \but as a single-line binary
based on elliptical-orbit solution for the \ha emission RVs -- cf.
Table~\ref{elem13}. The estimates are derived assuming the primary mass
of $M_1=2.9$~\ms, $A$ and $A_{\rm peri.}$ denote the semi-major axis
and the binary separation at periastron, respectively.}
\label{binary}
\begin{tabular}{lllcc}
\hline\noalign{\smallskip}
$i$  & $M_2/M_1$  & $M_2$ & $A$ & $A_{\rm peri.}$ \\
($^\circ$)  &     & (\ms) & (\rs) & (\rs)\\
\noalign{\smallskip}\hline
\hline\noalign{\smallskip}
90&0.0876&0.254&223.5&53.0\\
70&0.0936&0.272&223.9&53.1\\
50&0.1164&0.338&225.5&53.4\\
\hline\noalign{\smallskip}
\end{tabular}
\end{flushleft}
\end{table}

The results of Table~\ref{binary} show that the binary properties,
especially the low mass ratio, are quite similar to other binaries discovered
so far with Be primaries. For the estimates, we only considered higher
orbital inclinations since \but is one of the cases of an inverse
correlation between the brightness and emission-line strength, which
indicates that we see the system roughly equator-on -- cf, e.g.,
\citet{hvar83}.

If we adopt the distance to Pleiades $d=138$~pc after \citet{gro2007},
we estimate that the projected angular distance of the binary components
should be $\theta=0\farcs0075$, dropping down to $0\farcs0018$ at
periastron. This angular separation is certainly within reach
of existing large optical interferometers. The only
problem is the luminosity ratio primary/secondary.
If the secondary would be a normal late M dwarf corresponding to its mass,
it would be fainter in the visual region by more than 10 magnitudes
and the only chance to search for it would be in the far IR region, where,
however, the IR excess from the Be envelope can complicate the
detection. However -- if it were
a hot subdwarf, similar to the one found for another Be binary $\varphi$~Per
by \citet{gies98} -- it might be observable in the optical region
since the absolute visual magnitude of \but is fainter for some 2 magnitudes
than for the $\varphi$~Per B0.5e primary. Finally, a cool
Roche-lobe filling secondary seems improbable since it would probably
produce binary eclipses.

In any case, attempts to resolve the 218-d binary system with some
large interferometer are very desirable since
a visual orbit would help not only to estimate the true orbital
inclination but also to clarify whether the orbit has a high
eccentricity or is nearly circular.

\section{Comments on Hirata's model}

\begin{figure}
\centering
\resizebox{\hsize}{!}{\includegraphics{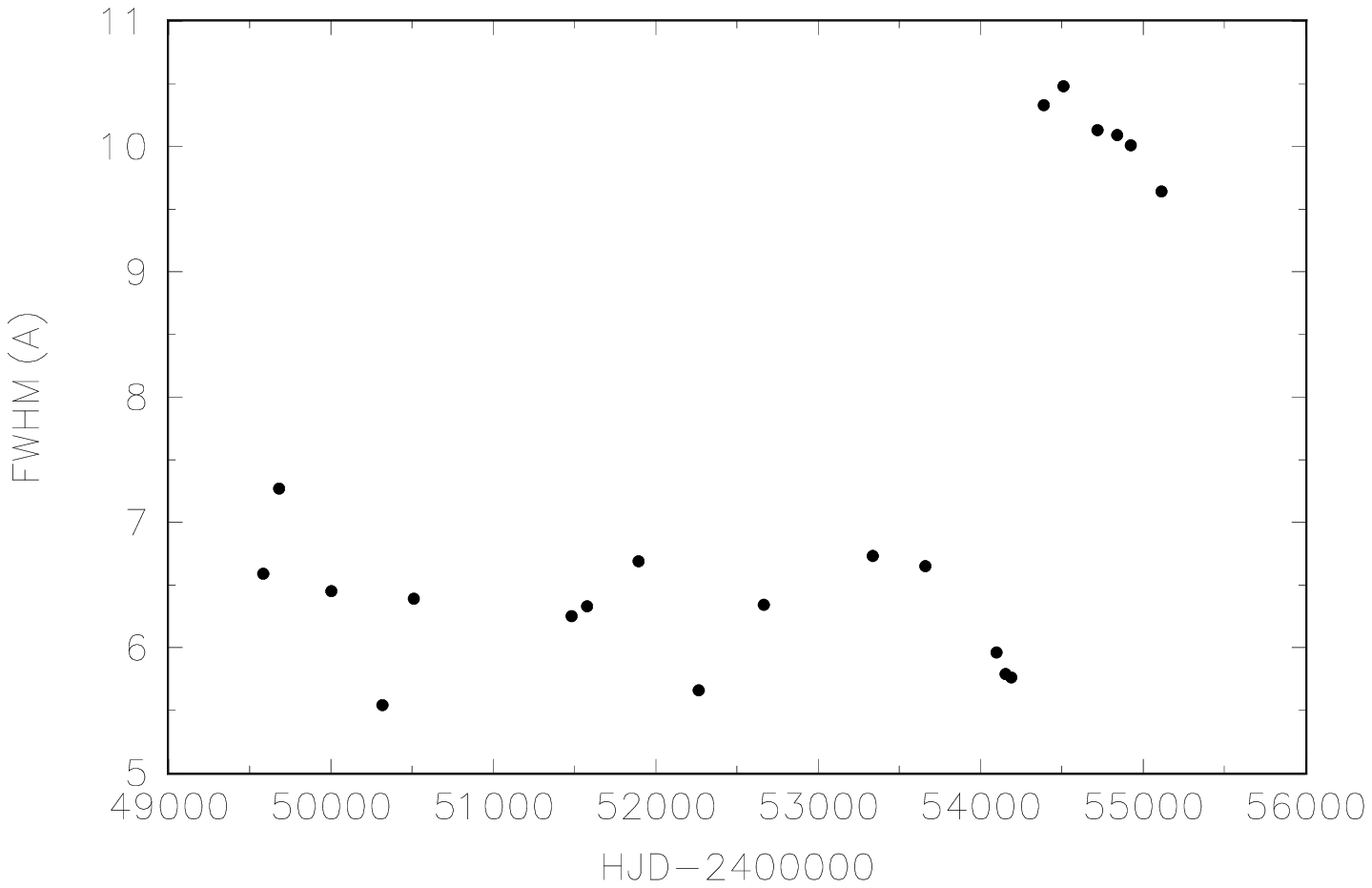}}
\caption{A time development of the FWHM (in \ANG) of the \ha emission.
The rapid increase is caused by the formation and a fast strengthening
of another double emission due to a newly formed envelope.}
\label{fwhm}
\end{figure}

\begin{figure}
\centering
\resizebox{\hsize}{!}{\includegraphics{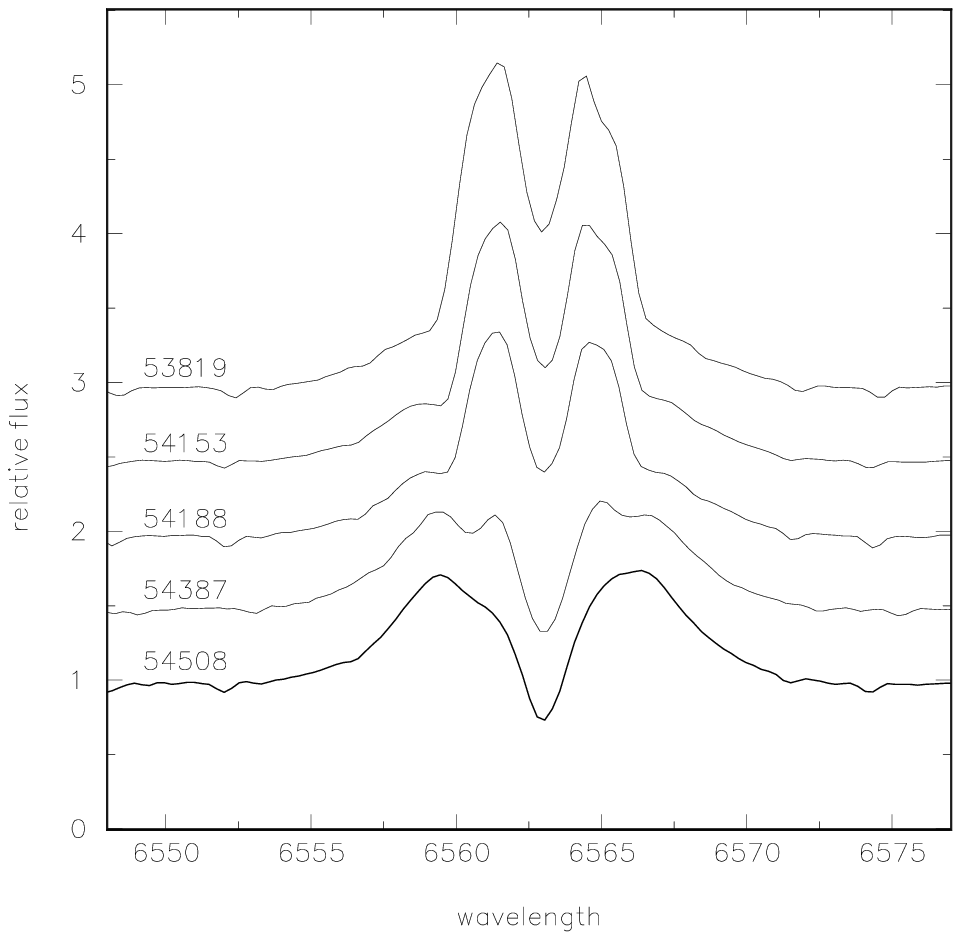}}
\caption{A series of the \ha profiles over the time interval of the
formation of a new shell. The HJDs-2400000 of individual spectra are
shown and the time runs from the top to the bottom. One can see how
the new broad emission gradually rises in intensity and how its blending
with the decaying previous double (but narrower) emission creates a
profile with four emission peaks for some time. Then the new
emission gets so strong that it merges with the original one.}
\label{shell}
\end{figure}

We have postponed a detailed study of the long-term changes for a
later work (Iliev et al. in prep.), but we wish to comment briefly on
the hypothesis put forward recently by \citet{hirata2007}.
He obtained systematic spectroscopy and polarimetry
of \but from 1974 to 2003 and finds a change in the polarization
angle from about 60$^\circ$ to 130$^\circ$ over that time interval.
He interprets this change as evidence of the precession of
the circumstellar disk that is responsible for the observed \ha emission.
He further argues that also the change in the \ha profiles from a weak
double emission with a strong central absorption core to a strong
emission with a wine-bottle shape indicates that the disk was first
seen more or less edge-on and later more face-on. \citet{tana2007}
studied the spectra of \but from Nov. 2005 until April 2007, which cover
the period of a formation of the new shell phase. They argue that
a new disk was formed in the equatorial plane of the B star while
the old disk was decaying but still present. According to their
interpretation, the old disk was precessing in space as suggested
by \citet{hirata2007}. Our spectra cover a much longer time interval,
including the one studied by \citet{tana2007}, and as Fig.~\ref{h3time}
shows, the change of the \ha profile was smooth. We thus measured
the full width at half maximum (FWHM) of a representative selection
of our \ha emission-line profiles and the variation in FWHM with time is shown
in Fig.~\ref{fwhm}. It was already demonstrated by \citet{struve31} in his
first model of Be stars as rapidly rotating objects that there is a clear
correlation between the width of presumably photospheric \ion{He}{I}
lines and the width of the Balmer emission lines, which is preserved
during the long-term changes. This correlation has been confirmed by a number
of later studies -- see, e.g., Fig.~5 of \citet{slet79}. One would
therefore expect that, if the appearance of a new shell phase
of \but is primarily a consequence of a geometrical effect,
namely a gradual precession of a flat disk that becomes
to be seen equator-on, the FWHM should gradually grow as
the new shell phase is approaching.
In contrast, Fig.~\ref{fwhm} shows that the FWHM of \ha {\sl was
 slowly decreasing} during the last 15 years. Its dramatic increase
is related to the formation of a new envelope, which our spectra
clearly confirm -- see Fig.~\ref{shell}. The apparent discontinous
increase in the FWHM occurs at the moment when the strength of the
broader emission from the new envelope rises to a half of the
peak intensity of the original emission. All this indicates that
the observed variations are primarily due to physical changes
in the circumstellar matter and cannot be reduced to a simple
geometrical cause -- a precession of the original gaseous disk.
There has been a rather widespread tendency in recent years to intepret
the presence of shell absorption lines as evidence of an equator-on view,
since many investigators are picturing the Be star disk as a flat structure
located at the stellar equator with a (rather small) opening angle
\citep{waters86,bjor93,hanus95,hanus96}. It is true that this model
can lead to theoretical Balmer profiles similar to the observed ones,
see, e.g., the 3D radiative line transfer models by \citet{hum94}.
One should be aware, however, that there is no unique proof of a
specific geometry on the level of various simplifications of current models.
For instance, H\"oflich~(1987,1988) succeeded in modeling several
Balmer emission-line profiles of particular Be stars with his
model consisting of an NLTE atmosphere and a {\sl spherical} envelope.
It is then conceivable that strong shell lines could also develop in
the spectrum of a Be star seen more or less pole-on in situations where
a very extended {\sl spheroidal envelope} forms around it.
Similarly, it might be worth considering whether the asymmetry detected
by the gradual change in the polarimetric angle is indeed caused by
the precession of a flat disk or by some other effect,
e.g. by a slowly revolving elongated (non-axisymmetric) disk.

\begin{acknowledgements}
We profited from the use of the program SPEL, written by our late
colleague Dr.~Ji\v{r}\'\i\ Horn.
We acknowledge the use of the publicly available Elodie spectra from the
electronic archive of the Haute Provence Observatory.
Our thanks go to Drs. M.~Ceniga, P.~Hadrava, A.~Kawka,
D.~Kor\v{c}\'akov\'a, J.~Krti\v{c}ka, M.~Netolick\'y, S.~\v{S}tefl, and
V.~Votruba, who secured some of the Ond\v{r}ejov spectrograms used
in this study. We also thank the referee, Dr. A.F.~Gulliver, for his
comments on the first version of the paper.
The research of the Czech authors was supported by the grant
205/06/0304 and 205/08/H005 of the Czech Science Foundation
and also from the Research Programs MSM0021620860
{\sl Physical study of objects and processes in the solar system
and in astrophysics} of the Ministry of Education of the Czech Republic,
and AV0Z10030501 of the Academy of Sciences of the Czech Republic.
The research of PK was supported by the ESA PECS grant 98058.
In its final stages, the research of JN, PH, and MW was also supported by
the grant P209/10/0715 of the Czech Science Foundation.
We acknowledge the use of the electronic database from the CDS, Strasbourg and
electronic bibliography maintained by the NASA/ADS system.
\end{acknowledgements}

\bibliographystyle{aa}
\bibliography{13885bib1}

\begin{thebibliography}{45}
\expandafter\ifx\csname natexlab\endcsname\relax\def\natexlab#1{#1}\fi

\bibitem[{{Ballereau} {et~al.}(1988){Ballereau}, {Chauville}, \&
  {Mekkas}}]{bal88}
{Ballereau}, D., {Chauville}, J., \& {Mekkas}, A. 1988, \aaps, 75, 139

\bibitem[{{Bjorkman} \& {Cassinelli}(1993)}]{bjor93}
{Bjorkman}, J.~E. \& {Cassinelli}, J.~P. 1993, \apj, 409, 429

\bibitem[{{Bo\v{z}i\'c} {et~al.}(1995){Bo\v{z}i\'c}, {Harmanec}, {Horn},
  {Koubsk\'y}, {Scholz}, {McDavid}, {Hubert}, \& {Hubert}}]{bozic95}
{Bo\v{z}i\'c}, H., {Harmanec}, P., {Horn}, J., {et~al.} 1995, \aap, 304, 235

\bibitem[{{Doazan} {et~al.}(1988){Doazan}, {Bourdonneau}, \&
  {Thomas}}]{doazan88}
{Doazan}, V., {Bourdonneau}, B., \& {Thomas}, R.~N. 1988, \aap, 205, L11

\bibitem[{{Gies} {et~al.}(1998){Gies}, {Bagnuolo}, {Ferrara}, {Kaye},
  {Thaller}, {Penny}, \& {Peters}}]{gies98}
{Gies}, D.~R., {Bagnuolo}, Jr., W.~G., {Ferrara}, E.~C., {et~al.} 1998, \apj,
  493, 440

\bibitem[{{Gies} {et~al.}(1990){Gies}, {McKibben}, {Kelton}, {Opal}, \&
  {Sawyer}}]{gies90}
{Gies}, D.~R., {McKibben}, W.~P., {Kelton}, P.~W., {Opal}, C.~B., \& {Sawyer},
  S. 1990, \aj, 100, 1601

\bibitem[{{Groenewegen} {et~al.}(2007){Groenewegen}, {Decin}, {Salaris}, \& {De
  Cat}}]{gro2007}
{Groenewegen}, M.~A.~T., {Decin}, L., {Salaris}, M., \& {De Cat}, P. 2007,
  \aap, 463, 579

\bibitem[{{Gulliver}(1977)}]{gul77}
{Gulliver}, A.~F. 1977, \apjs, 35, 441

\bibitem[{{Hanuschik}(1995)}]{hanus95}
{Hanuschik}, R.~W. 1995, Be Star Newsletter, 30, 17

\bibitem[{{Hanuschik}(1996)}]{hanus96}
{Hanuschik}, R.~W. 1996, \aap, 308, 170

\bibitem[{{Harmanec}(1982)}]{bebin82}
{Harmanec}, P. 1982, in IAU Symp. 98: Be Stars, 279--293

\bibitem[{{Harmanec}(1983)}]{hvar83}
{Harmanec}, P. 1983, Hvar Observatory Bulletin, 7, 55

\bibitem[{{Harmanec}(1988)}]{mr88}
{Harmanec}, P. 1988, Bulletin of the Astronomical Institutes of Czechoslovakia,
  39, 329

\bibitem[{{Harmanec} {et~al.}(2000){Harmanec}, {Habuda}, {{\v S}tefl},
  {Hadrava}, {Kor{\v c}{\'a}kov{\'a}}, {Koubsk{\'y}}, {Krti{\v c}ka},
  {Kub{\'a}t}, {{\v S}koda}, {{\v S}lechta}, \& {Wolf}}]{zarf20}
{Harmanec}, P., {Habuda}, P., {{\v S}tefl}, S., {et~al.} 2000, \aap, 364, L85

\bibitem[{{Hirata}(1995)}]{hirata95}
{Hirata}, R. 1995, \pasj, 47, 195

\bibitem[{{Hirata}(2007)}]{hirata2007}
{Hirata}, R. 2007, in Astronomical Society of the Pacific Conference Series,
  Vol. 361, Active OB-Stars: Laboratories for Stellare and Circumstellar
  Physics, ed. {A.~T.~Okazaki, S.~P.~Owocki, \& S.~Stefl}, 267--271

\bibitem[{{Hirata} \& {Kogure}(1976)}]{hirata76}
{Hirata}, R. \& {Kogure}, T. 1976, \pasj, 28, 509

\bibitem[{{Hirata} \& {Kogure}(1977)}]{hirata77}
{Hirata}, R. \& {Kogure}, T. 1977, \pasj, 29, 477

\bibitem[{{Horn} {et~al.}(1996){Horn}, {Kub\'at}, {Harmanec}, {Koubsk\'y},
  {Hadrava}, {\v{S}imon}, {\v{S}tefl}, \& {\v{S}koda}}]{sef0}
{Horn}, J., {Kub\'at}, J., {Harmanec}, P., {et~al.} 1996, \aap, 309, 521

\bibitem[{{Hummel}(1994)}]{hum94}
{Hummel}, W. 1994, \aap, 289, 458

\bibitem[{{Iliev} {et~al.}(2007){Iliev}, {Koubsk{\'y}}, {Kub{\'a}t}, \&
  {Kawka}}]{ili2007}
{Iliev}, L., {Koubsk{\'y}}, P., {Kub{\'a}t}, J., \& {Kawka}, A. 2007, in
  Astronomical Society of the Pacific Conference Series, Vol. 361, Active
  OB-Stars: Laboratories for Stellare and Circumstellar Physics, ed. A.~T.
  {Okazaki}, S.~P. {Owocki}, \& S.~{Stefl}, 440--442

\bibitem[{{Iliev} {et~al.}(1988){Iliev}, {Kovachev}, \& {Ruusalepp}}]{ili88}
{Iliev}, L., {Kovachev}, B., \& {Ruusalepp}, M. 1988, Information Bulletin on
  Variable Stars, 3204, 1

\bibitem[{{Katahira} {et~al.}(1996{\natexlab{a}}){Katahira}, {Hirata}, {Ito},
  {Katoh}, {Ballereau}, \& {Chauville}}]{kat96a}
{Katahira}, J.-I., {Hirata}, R., {Ito}, M., {et~al.} 1996{\natexlab{a}}, in
  Revista Mexicana de Astronomia y Astrofisica, vol. 27, Vol.~5, Revista
  Mexicana de Astronomia y Astrofisica Conference Series, ed. V.~{Niemela},
  N.~{Morrell}, P.~{Pismis}, \& S.~{Torres-Peimbert}, 114--116

\bibitem[{{Katahira} {et~al.}(1996{\natexlab{b}}){Katahira}, {Hirata}, {Ito},
  {Katoh}, {Ballereau}, \& {Chauville}}]{kat96b}
{Katahira}, J.-I., {Hirata}, R., {Ito}, M., {et~al.} 1996{\natexlab{b}}, \pasj,
  48, 317

\bibitem[{{Koubsk{\'y}} {et~al.}(2000){Koubsk{\'y}}, {Harmanec}, {Hubert},
  {Floquet}, {Kub{\'a}t}, {Ballereau}, {Chauville}, {Bo{\v z}i{\'c}},
  {Holmgren}, {Yang}, {Cao}, {Eenens}, {Huang}, \& {Percy}}]{zarf19}
{Koubsk{\'y}}, P., {Harmanec}, P., {Hubert}, A.~M., {et~al.} 2000, \aap, 356,
  913

\bibitem[{{Luthardt} \& {Menchenkova}(1994)}]{luth94}
{Luthardt}, R. \& {Menchenkova}, E.~V. 1994, \aap, 284, 118

\bibitem[{{McAlister} {et~al.}(1989){McAlister}, {Hartkopf}, {Sowell},
  {Dombrowski}, \& {Franz}}]{mcal89}
{McAlister}, H.~A., {Hartkopf}, W.~I., {Sowell}, J.~R., {Dombrowski}, E.~G., \&
  {Franz}, O.~G. 1989, \aj, 97, 510

\bibitem[{{Merrill}(1952)}]{mer52}
{Merrill}, P.~W. 1952, \apj, 115, 145

\bibitem[{{Miroshnichenko} {et~al.}(2002){Miroshnichenko}, {Bjorkman}, \&
  {Krugov}}]{miro2002}
{Miroshnichenko}, A.~S., {Bjorkman}, K.~S., \& {Krugov}, V.~D. 2002, \pasp,
  114, 1226

\bibitem[{{Miroshnichenko} {et~al.}(2001){Miroshnichenko}, {Fabregat},
  {Bjorkman}, {Knauth}, {Morrison}, {Tarasov}, {Reig}, {Negueruela}, \&
  {Blay}}]{miro2001}
{Miroshnichenko}, A.~S., {Fabregat}, J., {Bjorkman}, K.~S., {et~al.} 2001,
  \aap, 377, 485

\bibitem[{{Moultaka} {et~al.}(2004){Moultaka}, {Ilovaisky}, {Prugniel}, \&
  {Soubiran}}]{moul2004}
{Moultaka}, J., {Ilovaisky}, S.~A., {Prugniel}, P., \& {Soubiran}, C. 2004,
  \pasp, 116, 693

\bibitem[{{Rivinius} {et~al.}(2006){Rivinius}, {{\v S}tefl}, \&
  {Baade}}]{rivi2006}
{Rivinius}, T., {{\v S}tefl}, S., \& {Baade}, D. 2006, \aap, 459, 137

\bibitem[{{Roberts} {et~al.}(2007){Roberts}, {Turner}, \& {ten
  Brummelaar}}]{rob2007}
{Roberts}, Jr., L.~C., {Turner}, N.~H., \& {ten Brummelaar}, T.~A. 2007, \aj,
  133, 545

\bibitem[{{Sharov} \& {Lyuty}(1976)}]{sharov76}
{Sharov}, A.~S. \& {Lyuty}, V.~M. 1976, in IAU Symposium, Vol.~70, Be and Shell
  Stars, ed. A.~{Slettebak}, 105--106

\bibitem[{{Sharov} \& {Lyutyj}(1992)}]{sharov92}
{Sharov}, A.~S. \& {Lyutyj}, V.~M. 1992, \azh, 69, 544

\bibitem[{{\v{S}koda}(1996)}]{spefo}
{\v{S}koda}, P. 1996, in ASP Conf. Ser. 101: Astronomical Data Analysis
  Software and Systems V, 187--189

\bibitem[{{Slettebak}(1979)}]{slet79}
{Slettebak}, A. 1979, Space Science Reviews, 23, 541

\bibitem[{{Stellingwerf}(1978)}]{stel78}
{Stellingwerf}, R.~F. 1978, \apj, 224, 953

\bibitem[{{Struve}(1931)}]{struve31}
{Struve}, O. 1931, ApJ, 73, 94

\bibitem[{{Struve} \& {Swings}(1943)}]{struve43}
{Struve}, O. \& {Swings}, P. 1943, \apj, 97, 426

\bibitem[{{Tanaka} {et~al.}(2007){Tanaka}, {Sadakane}, {Narusawa}, {Naito},
  {Kambe}, {Katahira}, \& {Hirata}}]{tana2007}
{Tanaka}, K., {Sadakane}, K., {Narusawa}, S.-Y., {et~al.} 2007, \pasj, 59, L35

\bibitem[{{Taranova} {et~al.}(2008){Taranova}, {Shenavrin}, \&
  {Nadjip}}]{tara2008}
{Taranova}, O., {Shenavrin}, V., \& {Nadjip}, A.~D. 2008, Peremennye Zvezdy
  Prilozhenie, 8, 6

\bibitem[{{Vondr{\'a}k}(1969)}]{vondrak69}
{Vondr{\'a}k}, J. 1969, Bull. Astron. Inst. Czechosl., 20, 349

\bibitem[{{Vondr{\'a}k}(1977)}]{vondrak77}
{Vondr{\'a}k}, J. 1977, Bull. Astron. Inst. Czechosl., 28, 84

\bibitem[{{Waters}(1986)}]{waters86}
{Waters}, L.~B.~F.~M. 1986, \aap, 162, 121

\end{thebibliography}

\Online
\begin{appendix}
\section{Overview of available spectroscopic observations}\label{spered}
Here, we provide some details on the spectra used in this study and listed
in Table~\ref{jou} and on their reduction:

\begin{enumerate}
\item {\sl Ond\v{r}ejov spectra:} All 101 electronic spectrograms were
obtained in the coud\'e focus of the 2.0-m reflector and have a linear
dispersion of 17.2~\Am and a 2-pixel resolution 12600 (11-12~\kms
per pixel). The first 35 spectra were taken with a Reticon~1872RF linear
detector and cover a spectral region from 6300 to 6730~\ANG.
Complete reductions of these spectrograms were carried out by JN with
the program SPEFO, written by the late Dr.~J.~Horn and further developed by
Dr.~P.~\v{S}koda and more recently by Mr.~J.~Krpata -- see \citet{sef0} and
\citet{spefo}.
The remaining spectra were secured with an SITe-5 $800\times2000$ CCD detector
and cover a slightly longer wavelength interval 6260--6760~\ANG.
Their initial reductions (bias subtraction, flatfielding, creation of 1-D
images, and wavelength calibration) were carried out by M\v{S} in IRAF.

\item {\sl DAO spectra:} These spectrograms were obtained in the coud\'e
focus of the 1.22-m reflector of the Dominion Astrophysical Observatory
by SY, who also carried out their initial reductions (bias subtraction,
flatfielding, and creation of 1-D images). Their wavelength calibration was
carried out by JN in SPEFO. The spectra were obtained
with the 32121H spectrograph with the IS32R image slicer. The
detectors were UBC-1 4096x200 CCD for data before May 2005 and
SITe-4 4096x2048 CCD for data after May 2005. They
cover a wavelength region from
6150 to 6750~\ANG, have a linear dispersion of 10~\Am and 2-pixel
resolution of 21700 ( $\sim7$~\kms per pixel).

\item {\sl OHP spectra:} The public ELODIE archive of the Haute Provence
Observatory \citep{moul2004} contains 30 spectra listed as \bue, but some
of them are actually spectra of 27~Tau. We were able to recover 21 usable
spectra. For the purpose of this study, we extracted, rectified, and measured
only the red parts of these spectrograms.

\begin{figure}
\centering
\resizebox{\hsize}{!}{\includegraphics{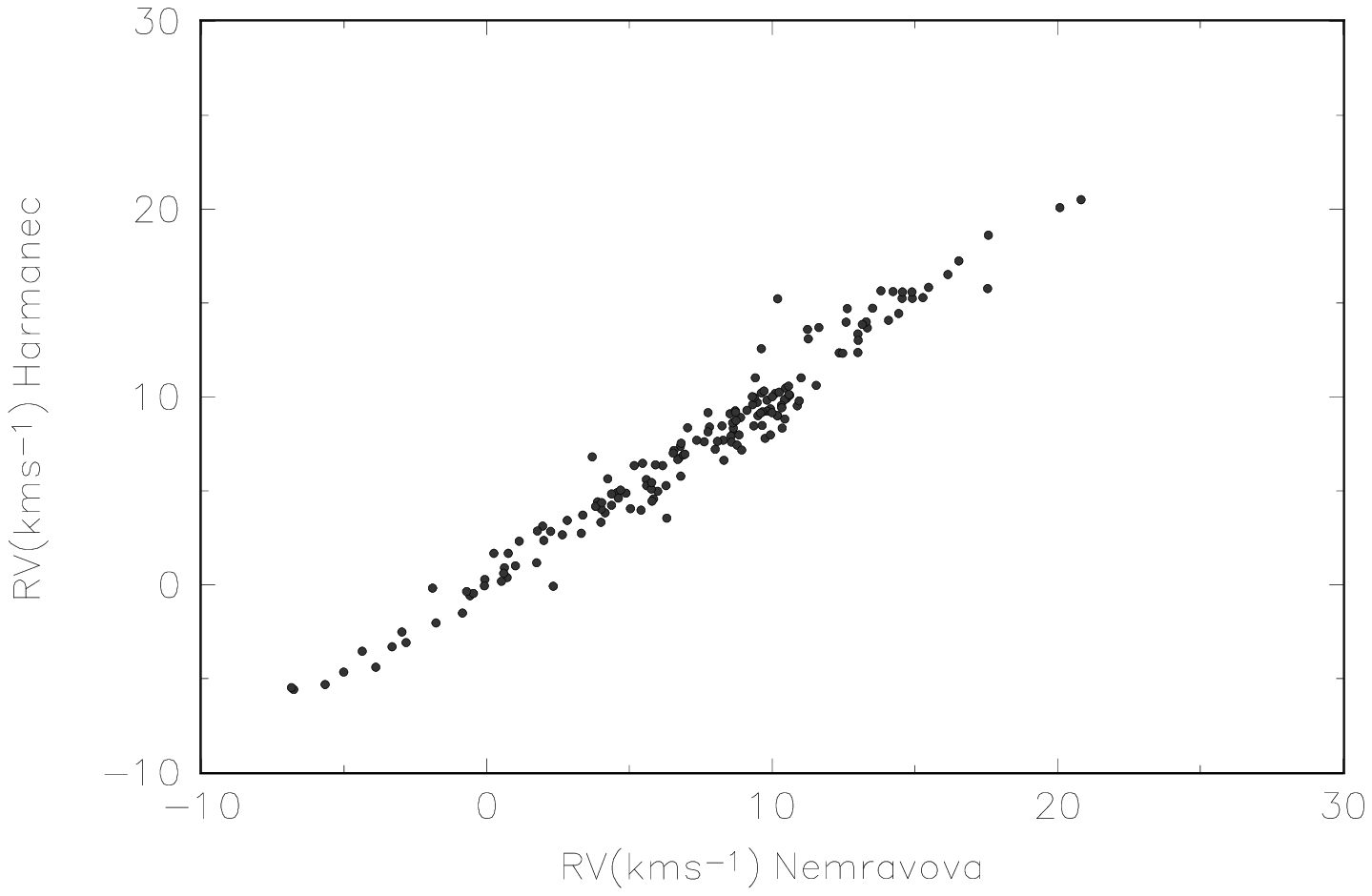}}
\resizebox{\hsize}{!}{\includegraphics{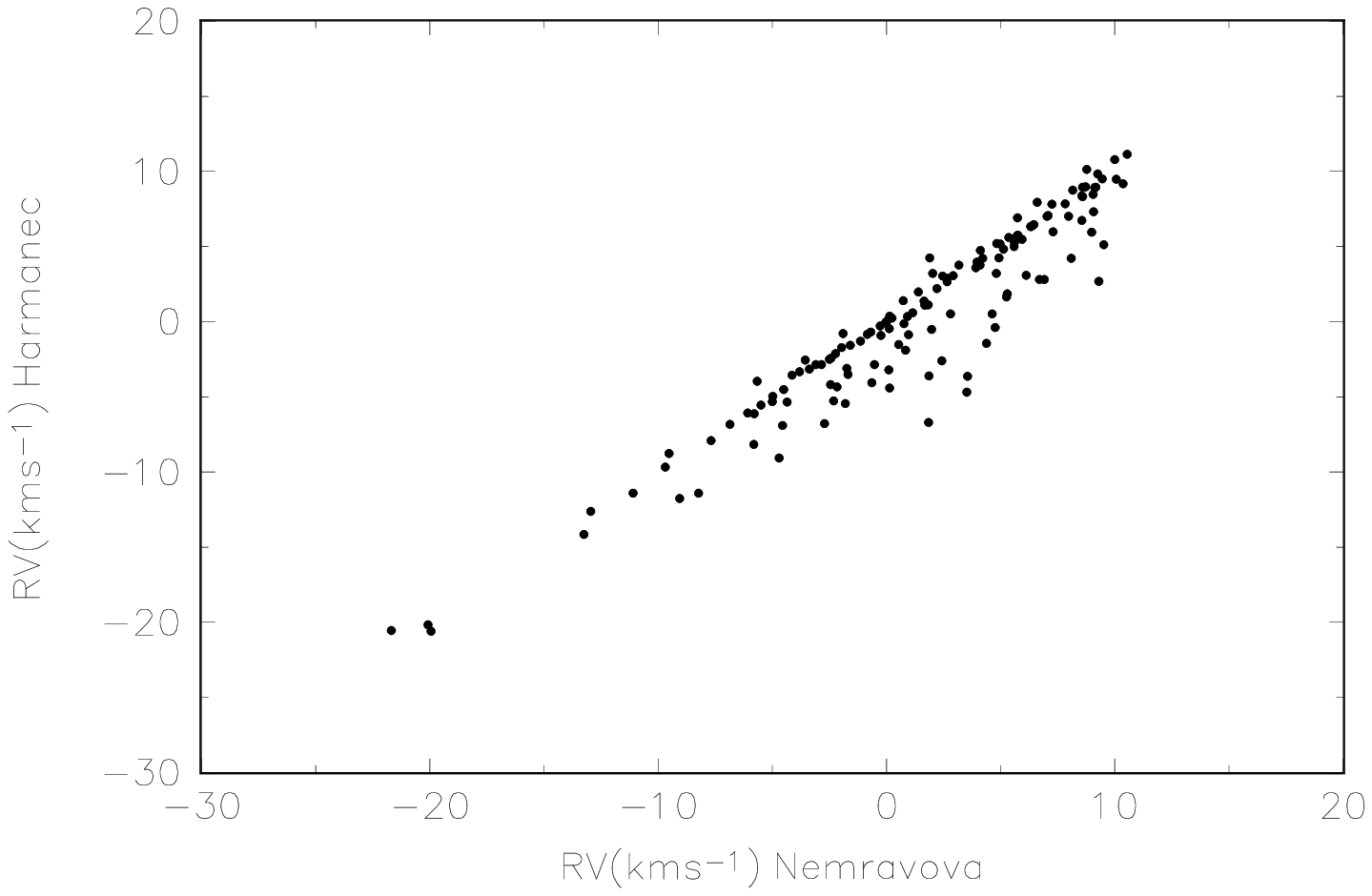}}
\caption{A comparison of independent RV measurements of the steep \ha
emission wings (upper panel) and shell core absorption (bottom panel).}
\label{rvkor}
\end{figure}

\item {\sl Rozhen spectra:} All 23 spectra from Rozhen observatory were
obtained in the coud\'e spectrograph of the 2-m RCC telescope. A CCD camera
Photometrics AT200 with SITe SI003AB~1024x1024 chip was used. The spectrograph
was used in a configuration providing high-resolution spectra suitable
for revealing fine details and the structure of the spectral lines.
A Bausch\&Lomb 632/22.3 grating was used in its 2nd order, giving
a linear dispersion of 4.2 A/mm with 2-pixel resolution of 33000
($\sim4.5$ km/s per pixel). Wavelength coverage is about 100~\ANG\ around \hae.
The initial reduction (bias subtraction, flatfielding, creation of 1-D
images and wavelength calibration) was carried out by LI in MIDAS.

\item {\sl Lisboa spectra:} These 4 CCD spectra were obtained with
the IGeoE 0.356-m SC telescope working at F/11. The spectrograph is
a Littrow LHIRESIII with a 2400~grooves per mm grating and a spectral
resolution of about 14.000. The initial reduction (bias subtraction,
flatfielding, creation of 1-D images, and wavelength calibration) of
the spectra was made by JR.
\end{enumerate}

The rectification and removal of cosmics and flaws of {\sl all spectrograms}
were carried out in a uniform way by JN in SPEFO.
The program SPEFO was also used to RV measurements, based on a comparison
of direct and flipped images of the spectral line profiles. Since we were
searching for small RV variations and since the setting on the steep wings
of the emission-line profiles was not always straightforward (see below),
these RV measurements were carried out independently by JN and PH. Besides
the settings on the steep wings of the \ha emission, we also measured
the \ha absorption core on all spectra where such absorption was present
to have a comparison with the results of \citet{kat96b}. We also tried
to measure RV of the \he absorption wings but due to weakness of this line
and its possible structure, these measurements turned out to be useless so
we did not use them. Following \citet{sef0}, we also measured selected
stronger and unblended telluric lines in all spectra and used them
to a correction of the RV zero point. Thanks to that, the spectra
from all observatories can be treated as coming from one instrument
for all practical purposes.

  A comparison of the two sets of independent RV measurements is shown in
Fig.~\ref{rvkor}. In general, the agreement is good.
A formal regression between the measurements of PH and JN
was derived. Its slope is $0.98 \pm 0.01$ for the \emi and
$0.94 \pm 0.01$ for the absorption. For the absorption
line, it is conceivable that in specific cases one or the other measurer
was confused by a telluric line blended with the stellar
absorption core. For analysis, we used the mean RVs of the two independent
measurements. All our RVs with the corresponding HJDs of their
mid-exposures are provided in Table~\ref{hjdrv}.

\begin{table*}
\begin{centering}
\caption{Radial velocities of the \ha emission wings and shell absorption
core obtained via averaging the independent measurements by
 J.~Nemravov\'a and P.~Harmanec; DAO = Dominion Astrophysical Observatory,
Victoria; ROZ = Rozhen National Observatory ; OND = Ond\v{r}ejov Observatory;
LIS = IGeoE-Lisbon; OHP = Haute Provence Observatory;
}\label{hjdrv}
\begin{tabular}{lrrllrrl}
\hline\noalign{\smallskip}
Time of obs.  & RV(\ha em.) & RV(\ha abs.) & Source & Time of obs. & RV(\ha em.) & RV(\ha abs.) & Source \\
(HJD-2400000) & [\kms]       & [\kms]      &        & (HJD-2400000) & [\kms]       & [\kms]      &        \\
\noalign{\smallskip}\hline
\hline\noalign{\smallskip}
49581.5875	&	4.94	&	--	&	OND	&	52957.5296	&	8.80	&	-0.28	&	OND	\\
49634.6281	&	0.76	&	--	&	OND	&	52978.2960	&	9.86	&	0.06	&	ROZ	\\
49644.5439	&	-3.96	&	--	&	OND	&	52992.4704	&	9.53	&	5.72	&	OND	\\
49658.4732	&	-0.54	&	--	&	OND	&	53027.3770	&	8.36	&	2.20	&	OND	\\
49659.4898	&	0.95	&	--	&	OND	&	53029.2971	&	9.82	&	3.96	&	OND	\\
49661.4455	&	1.21	&	--	&	OND	&	53042.2717	&	8.90	&	1.48	&	ROZ	\\
49662.5208	&	0.34	&	--	&	OND	&	53042.2776	&	8.73	&	0.74	&	ROZ	\\
49679.3450	&	2.32	&	-10.53	&	OND	&	53044.2905	&	10.57	&	4.86	&	ROZ	\\
49786.7179	&	2.17	&	-0.69	&	DAO	&	53044.3017	&	10.35	&	4.64	&	ROZ	\\
49930.5573	&	7.16	&	--	&	OND	&	53046.3601	&	9.36	&	3.09	&	ROZ	\\
49948.5677	&	9.24	&	--	&	OND	&	53048.3209	&	11.01	&	6.33	&	OND	\\
49949.6057	&	8.98	&	--	&	OND	&	53060.2802	&	9.59	&	5.48	&	OND	\\
50001.5337	&	9.63	&	--	&	OND	&	53082.2765	&	10.21	&	8.46	&	OND	\\
50015.4509	&	8.96	&	--	&	OND	&	53103.2588	&	9.65	&	2.68	&	ROZ	\\
50104.4137	&	4.68	&	--	&	OND	&	53216.5807	&	7.69	&	2.99	&	OND	\\
50122.3298	&	8.05	&	--	&	OND	&	53216.5841	&	6.72	&	2.61	&	OND	\\
50159.3244	&	7.52	&	--	&	OND	&	53236.5388	&	7.07	&	0.62	&	OND	\\
50316.5313	&	12.67	&	--	&	OND	&	53236.5418	&	6.67	&	1.69	&	OND	\\
50410.5517	&	12.17	&	--	&	OND	&	53236.5444	&	6.85	&	2.75	&	OND	\\
50439.3454	&	11.07	&	--	&	OND	&	53244.5837	&	7.85	&	0.32	&	ROZ	\\
50448.4556	&	12.39	&	--	&	OND	&	53303.4677	&	6.89	&	-3.16	&	ROZ	\\
50508.3366	&	9.58	&	--	&	OND	&	53303.4734	&	6.76	&	-2.61	&	ROZ	\\
50509.3213	&	8.40	&	--	&	OND	&	53306.4628	&	6.94	&	-2.43	&	ROZ	\\
51227.6834	&	15.86	&	--	&	DAO	&	53332.2934	&	6.14	&	-3.80	&	ROZ	\\
51481.5805	&	16.66	&	--	&	OND	&	53335.5298	&	4.87	&	-4.51	&	OND	\\
51570.2954	&	18.09	&	-11.26	&	OHP	&	53335.5058	&	4.15	&	-5.53	&	OND	\\
51572.2713	&	15.23	&	--	&	OHP	&	53452.2425	&	4.60	&	-5.45	&	ROZ	\\
51572.2823	&	15.27	&	-7.80	&	OHP	&	53555.5651	&	4.25	&	-6.60	&	OND	\\
51573.3101	&	14.43	&	--	&	OHP	&	53579.5663	&	4.92	&	-9.15	&	OND	\\
51573.3184	&	14.89	&	--	&	OHP	&	53615.5628	&	2.54	&	-6.85	&	OND	\\
51576.2801	&	20.66	&	--	&	OND	&	53628.0077	&	3.98	&	-4.82	&	DAO	\\
51797.5902	&	14.92	&	-3.58	&	OND	&	53638.9848	&	4.01	&	-5.96	&	DAO	\\
51888.4738	&	15.65	&	-12.79	&	OHP	&	53651.5938	&	2.64	&	-4.98	&	OND	\\
51889.4120	&	16.89	&	-1.85	&	OHP	&	53658.5135	&	3.02	&	-6.07	&	OND	\\
51892.4705	&	16.33	&	-3.26	&	OHP	&	53708.8482	&	3.65	&	-2.43	&	DAO	\\
51975.3321	&	14.73	&	-2.98	&	OND	&	53708.8552	&	3.68	&	-4.75	&	DAO	\\
52236.4148	&	12.34	&	-0.05	&	OHP	&	53745.4308	&	3.12	&	-3.33	&	OND	\\
52236.4313	&	13.49	&	-0.60	&	OHP	&	53745.4444	&	5.25	&	-3.26	&	OND	\\
52237.4193	&	13.64	&	1.47	&	OHP	&	53782.3483	&	-6.17	&	-13.71	&	ROZ	\\
52238.4212	&	15.07	&	-2.16	&	OHP	&	53791.3420	&	-6.16	&	-9.69	&	OND	\\
52239.4098	&	14.07	&	5.44	&	OHP	&	53791.6955	&	-4.84	&	-10.42	&	DAO	\\
52241.4048	&	13.66	&	2.18	&	OHP	&	53813.6682	&	0.54	&	-6.90	&	DAO	\\
52241.4168	&	13.28	&	-0.09	&	OHP	&	53813.6731	&	0.10	&	-7.00	&	DAO	\\
52242.3968	&	13.50	&	-2.44	&	OHP	&	53814.3319	&	-2.30	&	-11.32	&	ROZ	\\
52242.4088	&	13.17	&	-1.56	&	OHP	&	53814.7058	&	0.35	&	-5.74	&	DAO	\\
52263.3571	&	4.86	&	-21.12	&	OHP	&	53819.2800	&	1.72	&	-3.85	&	OND	\\
52264.3598	&	3.99	&	-20.28	&	OHP	&	54002.0002	&	-5.50	&	-5.17	&	DAO	\\
52264.3759	&	4.19	&	-20.14	&	OHP	&	54049.4053	&	1.00	&	-0.49	&	ROZ	\\
52287.7753	&	12.91	&	-3.06	&	DAO	&	54049.5688	&	0.59	&	-4.32	&	ROZ	\\
52533.0560	&	12.70	&	-0.59	&	DAO	&	54051.3857	&	-0.08	&	-0.54	&	ROZ	\\
52664.2587	&	12.66	&	6.15	&	OHP	&	54085.1936	&	4.30	&	6.64	&	OND	\\
52706.7375	&	9.20	&	3.07	&	DAO	&	54097.3352	&	5.58	&	5.29	&	OND	\\
52710.2434	&	8.94	&	-0.88	&	ROZ	&	54105.7532	&	6.28	&	4.59	&	DAO	\\
52860.5294	&	13.01	&	5.71	&	ROZ	&	54108.3564	&	4.60	&	3.58	&	ROZ	\\
52877.5835	&	14.12	&	4.46	&	OND	&	54115.3230	&	9.59	&	7.83	&	OND	\\
52899.5953	&	9.18	&	2.73	&	OND	&	54115.3365	&	6.24	&	5.07	&	OND	\\
52900.5672	&	10.16	&	1.06	&	OND	&	54116.3092	&	7.94	&	5.01	&	OND	\\
52900.5694	&	9.91	&	3.46	&	OND	&	54117.3291	&	8.49	&	4.97	&	OND	\\
52902.4322	&	11.08	&	0.23	&	OND	&	54126.2227	&	5.21	&	3.45	&	OND	\\
52902.4376	&	10.01	&	-0.84	&	OND	&	54153.2844	&	10.13	&	9.55	&	OND	\\
52904.6492	&	10.21	&	-1.22	&	OND	&	54162.3401	&	10.00	&	7.07	&	OND	\\
52904.6537	&	8.10	&	-1.58	&	OND	&	54164.2850	&	9.44	&	8.85	&	OND	\\
52904.6581	&	8.64	&	-2.50	&	OND	&	54186.2921	&	8.34	&	10.40	&	OND	\\
52949.6009	&	4.53	&	-4.85	&	OND	&	54188.3212	&	8.24	&	7.65	&	OND	\\
52949.6044	&	7.99	&	-1.69	&	OND	&	54341.0107	&	8.10	&	9.45	&	DAO	\\
52949.6091	&	8.48	&	0.25	&	OND	&	54387.5177	&	7.60	&	9.07	&	OND	\\
52952.5541	&	9.42	&	1.66	&	ROZ	&	54442.9976	&	-1.19	&	7.27	&	DAO	\\
52952.5619	&	8.61	&	0.73	&	ROZ	&	54490.2613	&	2.53	&	5.75	&	OND	\\
52957.5042	&	10.46	&	1.38	&	OND	&	54490.9272	&	5.12	&	7.32	&	DAO	\\
52957.5113	&	9.64	&	-0.03	&	OND	&	54508.3482	&	5.75	&	6.33	&	OND	\\
\hline
\end{tabular}
\end{centering}
\end{table*}

\setcounter{table}{1}
\begin{table*}
\begin{centering}
\caption{(cont.) Radial velocities of the \ha emission wings and shell
absorption core obtained via averaging the independent measurements by
 J.~Nemravov\'a and P.~Harmanec; DAO = Dominion Astrophysical Observatory,
Victoria; ROZ = Rozhen National Observatory ; OND = Ond\v{r}ejov Observatory;
LIS = IGeoE-Lisbon; OHP = Haute Provence Observatory;
}
\begin{tabular}{lrrllrrl}
\hline\noalign{\smallskip}
Time of obs. & RV(\ha em.) & RV(\ha abs.) & Source & Time of obs. & RV(\ha em.) & RV(\ha abs.) & Source \\
(HJD-2400000) & [\kms]       & [\kms]      &        & (HJD-2400000) & [\kms]       & [\kms]      &        \\
\noalign{\smallskip}\hline
\hline\noalign{\smallskip}
54519.6647	&	7.47	&	7.46	&	DAO	&	54871.4085	&	1.45	&	0.87	&	OND	\\
54537.3145	&	7.61	&	7.02	&	OND	&	54871.4347	&	-1.05	&	-1.35	&	OND	\\
54557.3011	&	9.58	&	7.53	&	OND	&	54872.2313	&	-0.59	&	0.00	&	OND	\\
54557.3179	&	9.66	&	10.86	&	OND	&	54872.2523	&	-0.47	&	-0.18	&	OND	\\
54718.5138	&	9.34	&	8.46	&	OND	&	54874.3436	&	-4.15	&	-2.85	&	LIS	\\
54748.4616	&	8.77	&	8.77	&	OND	&	54880.3360	&	-1.91	&	2.66	&	LIS	\\
54753.4457	&	10.20	&	9.03	&	OND	&	54881.3230	&	-2.97	&	2.77	&	LIS	\\
54753.4540	&	9.06	&	8.76	&	OND	&	54882.3222	&	-3.32	&	1.51	&	LIS	\\
54761.4009	&	10.36	&	9.77	&	OND	&	54911.6519	&	5.48	&	6.32	&	DAO	\\
54763.4827	&	10.24	&	8.19	&	OND	&	54911.6904	&	5.42	&	6.43	&	DAO	\\
54798.4322	&	10.35	&	9.47	&	OND	&	54912.6793	&	5.60	&	6.45	&	DAO	\\
54804.2374	&	9.88	&	8.46	&	OND	&	54924.3033	&	7.18	&	7.77	&	OND	\\
54840.4814	&	8.89	&	9.77	&	OND	&	55050.5248	&	11.68	&	8.25	&	OND	\\
54857.4154	&	8.08	&	7.49	&	OND	&	55071.5298	&	11.44	&	7.33	&	OND	\\
54862.6448	&	5.96	&	3.92	&	DAO	&	55083.6501	&	6.07	&	3.72	&	OND	\\
54862.6812	&	5.78	&	3.76	&	DAO	&	55097.4911	&	-1.45	&	4.41	&	OND	\\
54863.6276	&	5.43	&	4.58	&	DAO	&	55112.4081	&	3.52	&	-3.81	&	OND	\\
54863.6627	&	4.72	&	4.21	&	DAO	\\	
\hline
\end{tabular}
\end{centering}
\end{table*}

\end{appendix}
\end{document}